\documentclass[iop]{emulateapj}
\usepackage{apjfonts}

\usepackage[tight]{subfigure}
\usepackage{amsmath}
\usepackage{graphicx}
\usepackage{wasysym}
\usepackage{color}
\usepackage{verbatim}

\newcommand{\msun}{\ensuremath{\mathrm{M}_\odot}}

\def\kms{km\,s$^{-1}$}

\def\OI{O\,{\sc i}}

\def\CaII{Ca\,{\sc ii}}

\def\FeII{Fe\,{\sc ii}}
\def\FeIII{Fe\,{\sc iii}}

\def\CoIII{Co\,{\sc iii}}

\def\CaII{Ca\,{\sc ii}}

\def\Nifs{$^{56}$Ni}
\def\Cofs{$^{56}$Co}

\def\dm15{$\Delta m_{15}(B)$}
\def\MCh{M$_\mathrm{Ch}$}
\def\lesssim{\mathrel{\hbox{\rlap{\hbox{\lower4pt\hbox{$\sim$}}}\hbox{$<$}}}}

\def\gtrsim{\mathrel{\hbox{\rlap{\hbox{\lower4pt\hbox{$\sim$}}}\hbox{$>$}}}}

\shorttitle{[\OI] in the nebular spectrum of a subluminous SN~Ia}
\shortauthors{S. Taubenberger}

\begin{document}

\title{[\OI] $\lambda\lambda6300,6364$ in the nebular spectrum of a subluminous
  Type Ia supernova\footnotemark[*]}
\footnotetext[*]{Based on observations at ESO Paranal under prog.-ID
  088.D-0184.}

\author
{
  S.~Taubenberger\altaffilmark{1}, M.~Kromer\altaffilmark{1},
  R.~Pakmor\altaffilmark{2}, G.~Pignata\altaffilmark{3},
  K.~Maeda\altaffilmark{4}, S.~Hachinger\altaffilmark{5},
  B.~Leibundgut\altaffilmark{6} \& W.~Hillebrandt\altaffilmark{1}
}

\altaffiltext{1}
{Max-Planck-Institut f\"ur Astrophysik,
  Karl-Schwarzschild-Str. 1, 
  85741 Garching, Germany} 
\altaffiltext{2}
{Heidelberger Institut f\"{u}r Theoretische Studien, 
  Schloss-Wolfs\-brunnen\-weg 35, 
  69118 Heidelberg, Germany} 
\altaffiltext{3}
{Departamento de Ciencias Fisicas, Universidad Andres Bello,
  Avda. Republica 252, 
  Santiago, Chile} 
\altaffiltext{4}
{Kavli Institute for the Physics and Mathematics of the Universe
  (WPI), Todai Institutes for Advanced Study, University of Tokyo, 
  5-1-5 Kashiwanoha, Kashiwa, Chiba 277-8583, Japan} 
\altaffiltext{5}
{Julius-Maximilians-Universit\"at W\"urzburg, 
  Emil-Fischer-Str. 31, 97074 W\"urzburg, Germany} 
\altaffiltext{6}
{European Southern Observatory,
  Karl-Schwarzschild-Str.~2, 
  85748 Garching, Germany}

\begin{abstract}
  In this letter a late-phase spectrum of SN~2010lp, a subluminous
  Type Ia supernova (SN~Ia), is presented and analysed. As in
  1991bg-like SNe~Ia at comparable epochs, the spectrum is
  characterised by relatively broad [\FeII] and [\CaII] emission
  lines. However, instead of narrow [\FeIII] and [\CoIII] lines that
  dominate the emission from the innermost regions of 1991bg-like SNe,
  SN~2010lp shows [\OI] $\lambda\lambda6300,6364$ emission, usually 
  associated with core-collapse SNe and never observed in a subluminous 
  thermonuclear explosion before.  The [\OI] feature has a complex
  profile with two strong, narrow emission peaks. This suggests oxygen
  to be distributed in a non-spherical region close to the centre of
  the ejecta, severely challenging most thermonuclear explosion models
  discussed in the literature. We conclude that given these
  constraints violent mergers are presently the most promising
  scenario to explain SN~2010lp.
\end{abstract}


\section{Introduction}
\label{Introduction}

Despite the great success of Type Ia supernovae (SNe~Ia) as
cosmological distance indicators \citep{riess1998a,perlmutter1999a}
and the agreement that they represent thermonuclear explosions of CO
white dwarfs (WDs) in binary systems
\citep[e.g.][]{bloom2012a,hillebrandt2013a}, the details of their
explosion are not yet fully understood. There is e.g. an ongoing
debate on the nature of the companion star. Single-degenerate
scenarios with slightly evolved main-sequence-star or red-giant
companions are supported by the detection of circumstellar material in
some SNe~Ia (e.g. \citealt{patat2007a, dilday2012a}, but see
\citealt{livio2003a} and \citealt{soker2013a}). However, the
non-detection of the companion star of the SN~2011fe precursor in
multi-wavelength pre-explosion observations \citep{li2011a,bloom2012a}
and the lack of emission from supersoft X-ray sources in elliptical
galaxies \citep{gilfanov2010a} indicate that at least a fair fraction
of all SNe~Ia are probably produced by double-degenerate systems where
the companion is another WD. Closely related to the question of the
progenitor system is that of the explosion mechanism. Once ignited,
the thermonuclear flame may propagate subsonically (deflagration),
supersonically (detonation), or undergo a deflagration-to-detonation
transition \citep[e.g.][]{hillebrandt2013a}.

A powerful way to distinguish between different progenitor scenarios
and explosion mechanisms is by identifying their characteristic
nucleosynthesis footprints in observed SNe~Ia, since the distribution
of nucleosynthesis products varies between models. In pure
deflagrations, for instance, the ejecta are thoroughly mixed owing to
turbulent burning \citep[e.g.][]{roepke2007b,ma2013a}. Freshly
synthesised iron-group elements (IGEs) are present even in the
outermost ejecta, while unburned material is mixed to the core. The
influence on the spectral evolution is significant: pure deflagrations
may explain the subclass of peculiar 2002cx-like SNe
\citep{jha2006a,foley2013a,kromer2013a}, but not the bulk of SNe~Ia. 
Pure detonations and delayed detonations produce stratified ejecta, in
better agreement with observations of normal SNe~Ia
\citep[e.g.][]{hoeflich1996a,kasen2009a,sim2010a}.

Spectra of SNe~Ia taken around peak brightness probe the outer layers
of the ejecta rich in intermediate-mass elements (IMEs) and unburned
material. Only at later phases do the ejecta become sufficiently
transparent to directly see emission from the IGE-rich core. Nebular
spectra of SNe~Ia have therefore been recognised as a highly useful
tool to study nucleosynthesis and geometry effects in SNe~Ia
\citep[e.g.][]{axelrod1980a,kozma2005a,maeda2010c},
with important consequences for the preferred explosion scenarios.

All known nebular spectra of SNe~Ia are dominated by IGEs, most
notably by forbidden emission lines of \FeII\ and \FeIII, consistent
with inner ejecta mainly composed of IGEs produced in explosive Si
burning. In some SN~Ia subclasses (subluminous 1991bg-like objects,
2002cx-like objects and some superluminous SNe~Ia, see
\citealt{mazzali1997a,jha2006a,taubenberger2013a}) prominent [\CaII]
$\lambda\lambda7291,7323$ has been detected, but this is most likely
an ionisation rather than an abundance effect, with Ca being less
highly ionised in the overall `cooler' ejecta of those objects. A
feature often searched for in nebular spectra of SNe~Ia is the [\OI] 
$\lambda\lambda6300,6364$ doublet. Its presence would reveal unburned 
material in the inner parts of the ejecta, ruling out most of the 
proposed explosion mechanisms. However, so far [\OI] emission has never 
been detected in a subluminous SN~Ia, and only in a single normal SN~Ia, 
the ancient SN~1937C \citep{minkowski1939a}.

In this letter we present a late-time spectrum of the subluminous
SN~Ia 2010lp that shows a prominent emission feature at $\sim$\,6300\,\AA, 
probably due to [\OI] $\lambda\lambda6300,6364$. We analyse the line 
profile to constrain the spatial distribution of the emitting material, 
and consider what an [\OI] emission implies in terms of explosion 
scenarios. Finally, we discuss under what circumstances unburned 
material in the centre of SN~Ia ejecta may or may not produce an 
observable signature in form of late-time [\OI] 
$\lambda\lambda6300,6364$ emission.

\section{SN 2010lp}
\label{2010lp}

SN~2010lp was discovered on \textsc{ut} 2010 December 29.16 by
\citet{cox2010a} in the course of the Puckett Observatory Supernova
Search at an apparent unfiltered magnitude of 16.7. It was located at
$\alpha = 02^\mathrm{h}54^\mathrm{m}03\fs50$, $\delta =
+02\degr57\arcmin43\farcs4$ (J2000), 13.2 arcsec east of the centre of
NGC~1137, at a heliocentric redshift $z=0.010$ \citep{huchra1999a}.
Based on a spectrum taken on \textsc{ut} 2010 December 30--31,
\citet{prieto2011a} classified SN~2010lp as a subluminous SN~Ia
resembling SN~2007on \citep{stritzinger2011a} around maximum light.

Given the peculiar nature of the SN, its relative proximity and the
availability of early-time follow-up observations (Pignata et al. in
prep.), we selected SN~2010lp as a target for our programme on nebular
SN~Ia spectroscopy with VLT + FORS2. Spectra were taken on \textsc{ut}
2011 September 27--29 (grism 300V) and \textsc{ut} 2011 September
29--30 (grism 300I) with a 1.0 arcsec slit, an atmospheric-dispersion 
corrector and exposure times of $3 \times 2700$\,s for each grism. 
They were pre-reduced following standard recipes within \textsc{iraf}, 
followed by an optimal, variance-weighted extraction \citep*{horne1986a}. 
The dispersion solution was established using arc-lamp exposures and 
checked against isolated nightsky lines. Observations of spectrophotometric 
standard stars obtained during the same nights were used to calibrate the
spectra in flux and to remove telluric absorptions. Given the late
phase of the SN (264\,d after maximum light; Pignata priv. comm.) no
relevant evolution is expected over the 3 days of our observations. We
therefore combined all spectra to increase the signal-to-noise ratio
and maximise the wavelength coverage.

\begin{figure}
  \centering 
  \includegraphics[width=\linewidth]{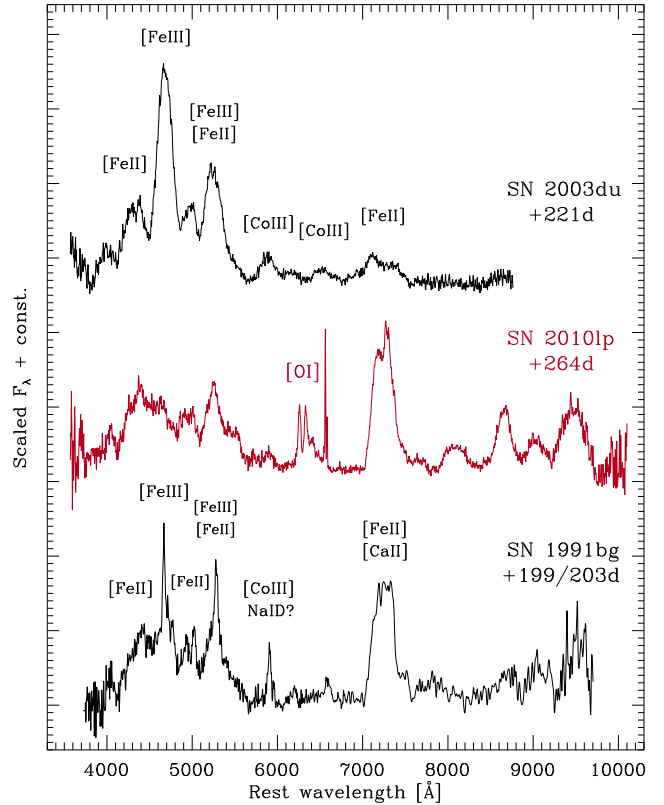}
  \caption{Nebular spectrum of SN~2010lp, compared with spectra of the
    normal SN~Ia 2003du \citep{stanishev2007b} and the subluminous
    SN~Ia 1991bg (\citealt{turatto1996a}; R.~Lopez priv. comm.) at
    similar epochs. The most important lines are identified.}
  \label{fig:spectrum}
\end{figure}

As in other SNe~Ia at late epochs, the resulting spectrum
(Fig.~\ref{fig:spectrum}) is dominated by forbidden emission lines of
Fe, but the ionisation state of the ejecta is low. [\FeIII] lines,
normally the hallmark features of nebular SN~Ia spectra, are weak or
absent, similar to what is observed in other subluminous SNe~Ia
\citep{mazzali1997a,mazzali2012a}. Instead, the blue part of the
spectrum up to $\sim$\,5500\,\AA\ is dominated by pseudo-continuous
emission from [\FeII] lines. A strong emission feature at
7000--7500\,\AA\ can be attributed to a combination of [\FeII] lines
(most notably [\FeII] $\lambda7155$) and [\CaII]
$\lambda\lambda7291,7323$. The strength of this feature is another
indication of a low ejecta ionisation, since in normal SNe~Ia calcium
is predominantly doubly ionised at those epochs.

Unprecedented in subluminous SNe~Ia is the emission feature seen 
at $\sim$\,6300\,\AA. In SNe~Ib/c, this feature is typically the most 
prominent emission line at nebular epochs, and regularly attributed 
to [\OI] $\lambda\lambda6300,6364$ \citep[e.g.][]{taubenberger2009a}.
The presence or absence of this feature has often been considered the 
sharpest observational discrimination between core-collapse and 
thermonuclear SNe \citep[e.g.][]{filippenko1997a}. In SN~2010lp, 
we observe it in a SN that is decidedly a SN~Ia as 
judged from its early-time spectrum \citep{prieto2011a} and the 
dominance of forbidden Fe lines in the nebular spectrum. We attribute 
the well-isolated emission (Fig.~\ref{fig:spectrum} -- compare with 
SN~1991bg) to [\OI] as in core-collapse SNe; IGEs, which are dominant 
in other parts of the spectrum, are -- owing to their complex level 
structure -- unlikely to produce single isolated features.

\section{Discussion}
\label{Discussion}

\subsection{[\OI] line profile}
\label{[OI] line profile}

Line profiles in nebular spectra are one-dimensional line-of-sight
projections of the three-dimensional emissivity distribution in the
ejecta. Analysing line profiles thus constrains the geometry of the
emission region, i.e. that part of the ejecta where 1) the emitting
chemical species is abundant, 2) has the correct ionisation state, and
3) the upper level of the respective transition is sufficiently
populated.  At the given epoch, the ionisation and excitation state
are largely determined by collisions of atoms with fast electrons and
positrons originating from the decay of radioactive elements such as
\Cofs.  This means that besides the oxygen distribution the spatial
distribution of \Nifs\,/\,\Cofs\ is relevant. For [\OI]
$\lambda\lambda6300,6364$ line-profile analyses are complicated by the
doublet nature of the line, with an intensity ratio depending on the
ambient matter density
\citep[e.g.][]{leibundgut1991a,spyromilio1991a}. In the following
analysis we assume the ratio of [\OI] $\lambda6300$ to [\OI]
$\lambda6364$ to be $\sim$\,3:1, appropriate for the optically thin
limit. This is supported by the non-detection of [\OI] $\lambda5577$,
which would be prominent if densities were higher. We fit the profile
of the [\OI] $\lambda\lambda6300,6364$ feature with multiple
components following \citet{taubenberger2009a}, where each component
actually consists of two Gaussian profiles with the same full width at
half maximum (FWHM), a fixed separation of 63.5\,\AA\ and an intensity
ratio of 3:1.

\begin{figure}
  \centering
  \includegraphics[width=\linewidth]{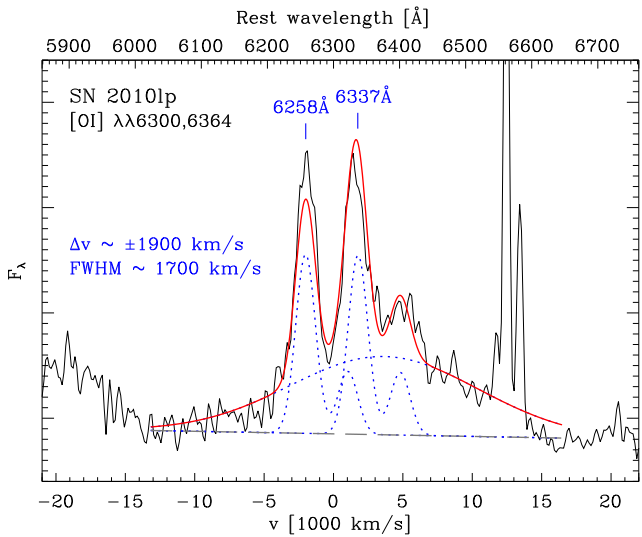}
  \caption{Profile of the [\OI] $\lambda\lambda6300,6364$ line in
    SN~2010lp, and its reproduction -- within the Gauss-fitting
    approach of \citet{taubenberger2009a} -- as a double-peaked
    structure atop a broad base. The narrow components have velocity
    offsets of $\sim$\,$\pm1900$ \kms\ from the rest position, and
    FWHM velocities of $\sim$\,1700 \kms.}
  \label{fig:line profile}
\end{figure}

\begin{figure}
  \centering
  \includegraphics[width=\linewidth]{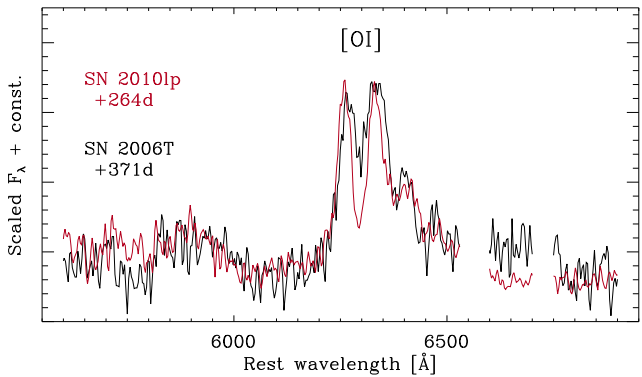}
  \caption{Comparison of the [\OI] $\lambda\lambda6300,6364$ line 
    profiles in SN~2010lp and the Type IIb SN~2006T \citep{maeda2008a}. 
    A geometric origin has been put forward for the [\OI] line profile 
    of SN~2006T \citep{maeda2008a,taubenberger2009a}.}
  \label{fig:CC}
\end{figure}

In SN~2010lp the [\OI] $\lambda\lambda6300,6364$ feature consists of
two similarly strong, fairly narrow emission peaks on top of a broad
base (Fig.~\ref{fig:line profile}). One of the narrow components 
is blueshifted with respect to the [\OI] $\lambda6300$ rest wavelength,
the other redshifted. Their separation is about 79\,\AA, too large for
them to originate from the two lines of the [\OI] doublet (with an
intensity ratio of $\sim$\,1:1 as expected in the optically thick limit). 
Instead, the observed line profile likely results from a complex 
geometry of the emission region, a conclusion also drawn for 
stripped-envelope core-collapse SNe with similar [\OI] profiles 
\citep[e.g.][see Fig.~\ref{fig:CC}]{maeda2008a,taubenberger2009a}.

In our Gauss-fitting approach we reproduce the profile with three
components (Fig.~\ref{fig:line profile}). The broad base has a FWHM
of $\sim$\,15\,000 \kms\ and is redshifted by $\sim$\,2900 \kms. It is
very likely a blend of [\OI] with other lines, possibly [\CoIII]. This
could explain both the unusual width and the apparent redshift. The
narrow components have FWHM velocities of $\sim$\,1700 \kms\ and
blue-\,/\,redshifts of 2000 and 1800 \kms, respectively. Geometrically
this could be interpreted as a large volume where some \OI\ is present
and emits at low intensity, complemented by a higher concentration of
\OI\ at low velocity responsible for the narrow emission features. To
obtain the observed double-peaked profile, the inner \OI\ emission
region cannot be spherically symmetric, but could e.g. consist of two
compact oxygen-rich blobs with opposite line-of-sight velocity, or a
torus-like structure viewed sideways
\citep{mazzali2005b,taubenberger2009a}.

\subsection{Comparison to other subluminous SNe Ia}
\label{Comparison to other subluminous SNe Ia}

\begin{figure*}
  \centering
  \includegraphics[width=0.72\linewidth]{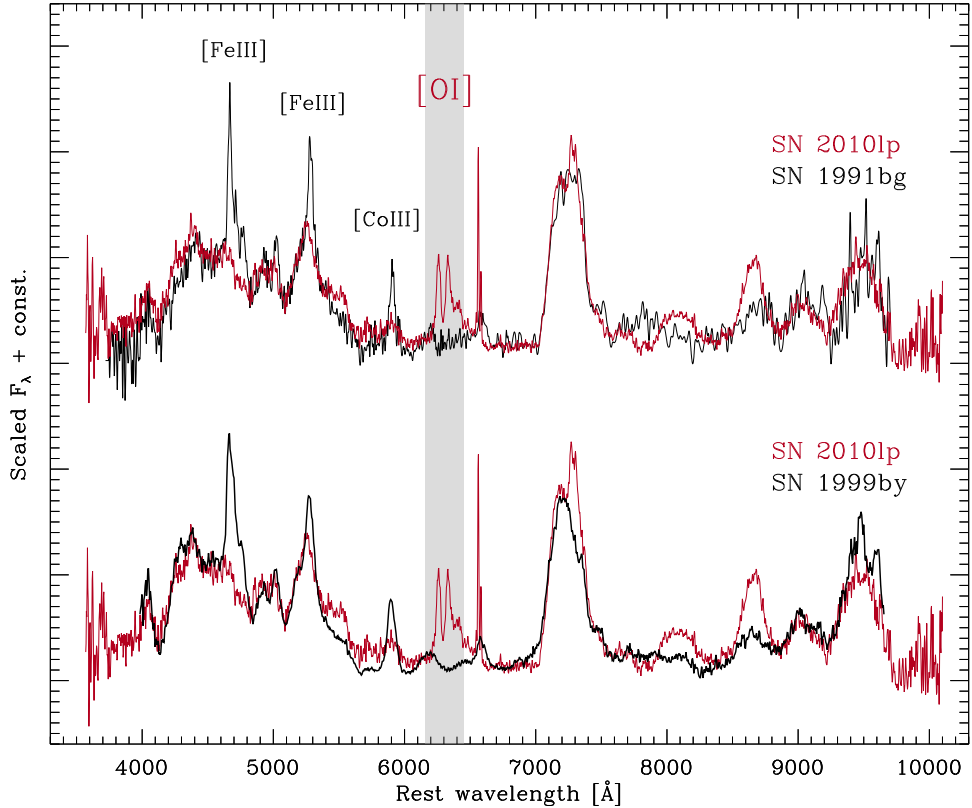}
  \caption{Spectrum of SN~2010lp (+264\,d) overplotted on spectra of
    the subluminous SNe~1991bg (+199\,/\,203\,d) and 1999by (+183\,d;
    \citealt{silverman2012a}). [\OI] $\lambda\lambda6300,6364$, which
    is clearly detected in SN~2010lp but absent in SNe~1991bg and
    1999by, is highlighted.}
  \label{fig:comparison}
\end{figure*}

A comparison of the nebular spectrum of SN~2010lp and those of the
subluminous SNe~1991bg and 1999by \citep{turatto1996a,silverman2012a}
is particularly insightful, since they share several peculiarities
distinguishing them from late-time spectra of normal SNe~Ia
(Figs.~\ref{fig:spectrum} and \ref{fig:comparison}). The strongest and
most characteristic features in nebular spectra of normal SNe~Ia,
i.e. broad (FWHM 7000--10\,000 \kms) [\FeIII] emission lines, are
absent in all subluminous objects. The pattern of emission features
arising from \FeII\ and \CaII, on the other hand, is remarkably
similar in SNe~1991bg, 1999by and 2010lp, suggesting that chemical
composition and ionisation state in these three SNe are almost
identical throughout most of the ejecta.

However, a significant difference is found in the emission arising
from the innermost zones. The spectra of `classical' subluminous
SNe~Ia show prominent narrow [\FeIII] and [\CoIII] emission lines
(FWHM $\sim$\,2000 and $\sim$\,3600 \kms\ in SNe~1991bg and 1999by,
respectively) which are entirely absent in SN~2010lp. In the latter,
in contrast, narrow [\OI] emission is detected with similar FWHM, not
present in SNe~1991bg and 1999by. However, what might appear to be a
simple `replacement' of iron by oxygen as the dominant coolant in the
innermost regions of SN~2010lp is actually more complicated. First of
all, the [\OI] emission in SN~2010lp is double-peaked, whereas the
[\FeIII] emission lines in SNe~1991bg and 1999by are single-peaked,
suggesting a different geometrical configuration of the emitting
material. Moreover, while the innermost ejecta in SNe~1991bg and
1999by must be more highly ionised than the surrounding material to
give rise to strong [\FeIII] lines [\citet{mazzali2012a} consider this
an indication for a low-density core], the ionisation in the
[\OI]-emitting region in SN~2010lp has to be rather low in order to
retain enough neutral oxygen.

\subsection{Confronting explosion models with SN 2010lp observations}
\label{Confronting explosion models with SN 2010lp observations}

To explain the narrow [\OI] emission detected in the nebular spectrum
of SN~2010lp is a severe challenge for SN~Ia explosion models. In
fact, the constraints on possible models are so stringent that most
scenarios discussed today can be rejected for SN~2010lp.

In pure deflagrations turbulent burning leads to thoroughly mixed
ejecta, with both IGEs and carbon\,/\,oxygen abundant at all
velocities \citep[e.g.][]{roepke2007b,ma2013a}. [\OI] emission in
nebular spectra might be expected in this scenario. However, given the
rather uniform distribution of unburned material, there is no reason
for the [\OI] emission to be mostly confined to a small region close
to the centre of the ejecta, as observed in SN~2010lp. Indeed
\citet{kozma2005a} found prominent \textit{broad} [\OI] emission in a
synthetic nebular spectrum for a pure-deflagration model. The thorough
mixing in deflagrations also enriches the outer ejecta layers with
IGEs compared to other scenarios. This results in peculiar early-time
spectra, long speculated \citep{jha2006a,phillips2007a} and recently
shown \citep{kromer2013a} to be similar to 2002cx-like SNe. Clearly,
this disagrees with the description of the early-phase spectrum of
SN~2010lp \citep{prieto2011a}. Taken together, pure deflagrations are
very unlikely to explain SN~2010lp.

In Chandrasekhar-mass (\MCh) delayed-detonation scenarios
\citep[e.g.][]{hoeflich1996a,kasen2009a,jordan2012a,seitenzahl2013a},
still considered to be favourable models for SNe~Ia, the centre is
entirely dominated by IGEs. [\OI] emission from the core is very
unlikely for these models. Similarly, no oxygen is expected to remain
in the central ejecta of most pure-detonation scenarios, including
sub-\MCh\ double detonations and edge-lit detonations
\citep[e.g.][]{nomoto1982b,livne1990a,fink2010a} as well as
spontaneous detonations in sub-\MCh\ WD-merger remnants
\citep{vankerkwijk2010a}. In all these scenarios the central density
is high enough for the detonation to convert the fuel almost
completely into IGEs.

There is probably only a single SN~Ia explosion channel discussed in
the literature that may yield oxygen in a narrow region close to the
centre of the ejecta: violent mergers \citep{pakmor2010a,pakmor2012a}.
In these models the primary WD ignites dynamically on the
surface. Depending on its mass, the emerging detonation burns most of
the primary to IGEs and IMEs. Once the secondary is hit by the ejecta,
it also ignites. However, since the density in the less massive and
tidally deformed secondary is lower, nuclear burning predominantly
proceeds to oxygen, which is ejected at relatively low velocity. If
the mass combination of the two WDs is sufficiently asymmetric as in
the $1.1+0.9$ \msun\ merger of \citet{pakmor2012a}, oxygen even fills
the innermost regions of the combined ejecta of the primary and
secondary, distributed aspherically (see Fig.~2 of
\citealt{pakmor2012a}).

Of course, the 1.1 + 0.9 \msun\ WD merger of \citet{pakmor2012a}
produces 0.61 \msun\ of \Nifs, and the early-time spectra do not
resemble subluminous SNe~Ia as reported for SN~2010lp. However, a
setup with a similar ratio of secondary to primary mass ($\sim$\,0.8), 
but a less massive primary, might produce sufficiently little \Nifs\ 
yet still leave oxygen close to the centre of the ejecta.

\subsection{Conditions for [\OI] formation}
\label{Conditions for [OI] formation}

In the qualitative analysis performed here we can only determine a
necessary condition for low-velocity [\OI] to be observed in nebular
spectra: the presence of neutral oxygen close to the core of the
ejecta. Whether this is sufficient cannot be assessed without full
NLTE modelling of the plasma state, since the formation of the line
may depend on details of the ejecta structure.

First of all, the relative location of oxygen and \Nifs\,/\,\Cofs\ is
crucial. A location sufficiently close to radioactive material helps
to populate the upper level of the [\OI] $\lambda\lambda6300,6364$
lines. However, too much heating by radioactive decay will ionise
oxygen too strongly. Moreover, if oxygen is mixed with IGEs on
microscopic scales, the ejecta might mostly cool in the numerous
forbidden lines of Fe rather than [\OI] $\lambda\lambda6300,6364$.

From these considerations it is clear that the sheer presence of
oxygen in the central part of the ejecta in violent-merger models may
not necessarily give rise to [\OI] $\lambda\lambda6300,6364$ emission
at late epochs. Therefore, SN~2010lp may not be unique in having a
core formed at least partially by oxygen. In more luminous SNe~Ia this
may also be present, but ionised owing to the more intense radiation
field. Hence, in spite of its oxygen core the 1.1 + 0.9 \msun\ WD
merger of \citet{pakmor2012a} could still be a viable model for
normally bright SNe~Ia, where nebular [\OI] emission has never been
observed. 1991bg-like SNe, however, have low \Nifs\ masses, and share
many similarities with SN~2010lp. To avoid [\OI] emission in those
objects, one cannot allow for much oxygen to be present near the
centre of the ejecta. This could be achieved if they were the outcome
of violent mergers of CO with He WDs, as proposed by
\citet{pakmor2013a}.

\section{Conclusions}
\label{Conclusions}

A late-phase spectrum of the subluminous Type Ia SN~2010lp, taken
$\sim$\,264\,d after maximum light, shares strong similarities with
spectra of 1991bg-like SNe~Ia at comparable epochs, being dominated by
emission of [\FeII] and [\CaII] lines. However, 1991bg-like SNe
additionally show narrow [\FeIII] and [\CoIII] lines, which are absent
in SN~2010lp. The latter instead shows a prominent feature near
6300\,\AA, which constitutes the first clear detection of [\OI]
$\lambda\lambda6300,6364$ in a nebular spectrum of a subluminous 
thermonuclear SN. Previously, [\OI] $\lambda\lambda6300,6364$ has been 
considered as the most direct observational evidence for a core-collapse 
origin of an explosion.

The profile of the [\OI] line is complex, formed by two narrow
emission peaks, one blue- and the other redshifted with respect to the
rest wavelength. The separation of these two features is too large to
be explained by the doublet nature of [\OI] $\lambda\lambda6300,6364$.
Therefore, we favour a geometric origin of the observed profile, with
a concentration of oxygen in a non-spherical region close to the
centre of the ejecta.

Most thermonuclear explosion models discussed in the literature do not
predict the presence of oxygen at low velocity and can therefore be
excluded for SN~2010lp. Pure-deflagration models do predict oxygen
throughout their strongly mixed ejecta, but no concentration in the
inner regions, so that the narrow emission peaks observed in SN~2010lp
cannot be explained. Moreover, pure deflagrations are ruled out from
the early spectra of SN~2010lp being similar to 1991bg-like SNe. In
some violent-merger models, however, oxygen is left near the centre of
the ejecta \citep{pakmor2011b,pakmor2012a}, making this the most
promising scenario for SN~2010lp currently available.

In the end, detailed 3D nebular spectrum synthesis calculations are needed
to better constrain the amount of oxygen, its spatial distribution and
the degree of mixing with \Nifs\ and its decay products necessary to
reproduce the complex [\OI] $\lambda\lambda6300,6364$ profile observed
in SN~2010lp. The ejecta structure obtained from hydrodynamic
simulations of violent mergers may serve as a starting point in such a
modelling approach.

\begin{acknowledgements}
  We thank David Branch and our referee, Saurabh Jha, for their 
  helpful comments. Comparison spectra in this paper were retrieved 
  from the \textsc{suspect} database
  (http://suspect.nhn.ou.edu/$\sim$suspect/). This work was supported
  by the Transregional Collaborative Research Centre TRR 33 of the DFG 
  and the Klaus Tschira Foundation. K.M. acknowledges support by the 
  WPI initiative, MEXT, Japan, G.P. by the Proyecto FONDECYT 11090421 
  and the Millennium Center for Supernova Science (P10-064-F).
\end{acknowledgements}

\bibliographystyle{apj}

\end{document}